\documentclass[aps,prl,twocolumn,superscriptaddress,amssymb,amsmath,nofootinbib]{revtex4-1}

\usepackage{amsmath,setspace,amsfonts,latexsym}
\usepackage{amssymb}
\usepackage{color}
\usepackage{epsfig}
\usepackage{graphicx}
\usepackage{slashed}
\usepackage{caption}
\usepackage{subcaption}

\definecolor{White}{rgb}{1,1,1}
\definecolor{Red}{rgb}{1,0.1,0}
\definecolor{LightYellow}{rgb}{1,1,.875}
\definecolor{SteelBlue}{rgb}{.273,.508,.703}
\definecolor{navy}{rgb}{0,0,.5}
\definecolor{LightCyan}{rgb}{.875,1,1}
\definecolor{DarkRed}{rgb}{.543,0,0}
\definecolor{HotPink}{rgb}{1,.41,.70}
\definecolor{ForestGreen}{rgb}{.13,.54,.13}
\definecolor{OliveDrab}{rgb}{.42,.55,.14}
\definecolor{MediumBlue}{rgb}{0,0,.80}
\definecolor{RoyalBlue}{rgb}{.25,.41,.88}
\definecolor{DeepSkyBlue}{rgb}{0,.746,1}
\definecolor{Brown}{rgb}{0.545,0.271,0.074}
\definecolor{Purple}{rgb}{0.637,0.285,0.641}

\def\bea{\begin{eqnarray}}
\def\eea{\end{eqnarray}}
\def\bec{\begin{center}}
\def\ec{\end{center}}

\def\beq{\begin{equation}}
\def\eeq{\end{equation}}

\newcommand\lsim{\mathrel{\rlap{\lower4pt\hbox{\hskip1pt$\sim$}}
    \raise1pt\hbox{$<$}}}
\newcommand\gsim{\mathrel{\rlap{\lower4pt\hbox{\hskip1pt$\sim$}}
    \raise1pt\hbox{$>$}}}
\def\bea{\begin{eqnarray}}
\def\eea{\end{eqnarray}}
\def\ba{\begin{array}}
\def\ea{\end{array}}
\def\bc{\begin{center}}
\def\ec{\end{center}}

\begin{document}

\title{\Large Glue to light signal of a new particle}

\author{Dongjin Chway}
\email{djchway@gmail.com}
\affiliation{Department of Physics and Astronomy
and Center for Theoretical Physics, Seoul National University, Seoul 151-747, Korea}
\author{Radovan Derm\'i\v{s}ek}
\email{dermisek@indiana.edu}
\affiliation{Department of Physics and Astronomy
and Center for Theoretical Physics, Seoul National University, Seoul 151-747, Korea}
\affiliation{Physics Department, Indiana University, Bloomington, IN 47405, USA}
\author{Tae Hyun Jung}
\email{thjung0720@gmail.com}
\affiliation{Department of Physics and Astronomy
and Center for Theoretical Physics, Seoul National University, Seoul 151-747, Korea}
\author{Hyung Do Kim}
\email{hdkim@phya.snu.ac.kr}
\affiliation{Department of Physics and Astronomy
and Center for Theoretical Physics, Seoul National University, Seoul 151-747, Korea}

\begin{abstract}

Any new particle charged under $SU(3)_C$ and carrying electric charge will leave an imprint in  the di-photon invariant mass spectrum as it can mediate $gg \to \gamma \gamma$ process through loops.
The combination of  properties of loop functions, threshold resummation  and gluon pdfs can result in a peak-like feature in the di-photon invariant mass around twice the mass of a given particle even if the particle is short-lived and thus it doesn't form a narrow bound state. 
Using recent ATLAS analysis, we set upper limits on the combined $SU(3)_C$ and electric charge of  new particles and indicate future prospects. 
We also discuss the possibility that  the excess of events in the di-photon invariant mass spectrum around 750 GeV originates  from loops of a particle with mass around 375 GeV.

\end{abstract}

\maketitle

\noindent
{\bf Introduction.}
As demonstrated by  discoveries of the Z boson and the Higgs boson, a resonance is the  cleanest signal of a new particle as long as its branching ratios to visible modes are nonzero. However, many popular models including minimal supersymmetric standard model or models with various top-partners predict particles that can be produced in  pairs. For the pair production, the searches highly depend on decay modes of a given particle and there are known scenarios in well motivated models  which are  difficult to see directly even if production cross sections are sizable. 
In principle, a model can always be constructed in which a new particle cascade decays to complex final states consisting of soft particles and possibly missing energy, or the particle has a large number of possible final states with small branching ratios to individual ones.
Signatures that are less model dependent or do not depend on decay modes at all are therefore an  integral part of  searches for new physics.

\begin{figure}[bht]
\begin{subfigure}[b]{0.4\textwidth}
\includegraphics[width=\textwidth]{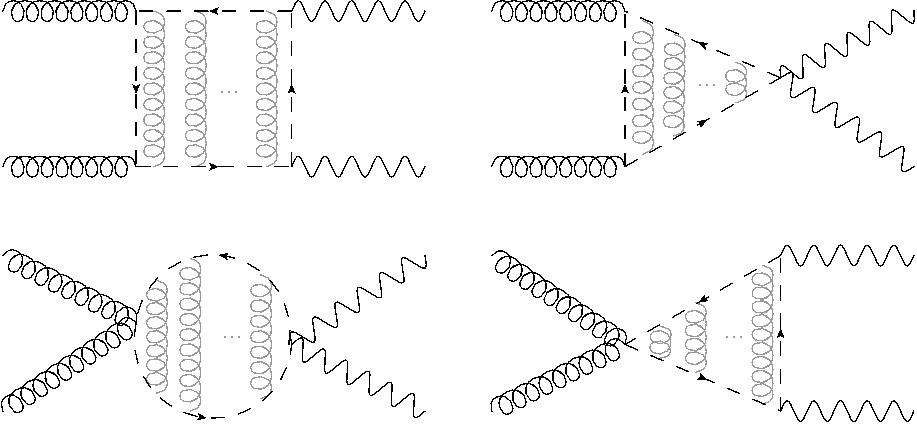}
\caption{}
\end{subfigure}
\begin{subfigure}[b]{0.2\textwidth}
\includegraphics[width=\textwidth]{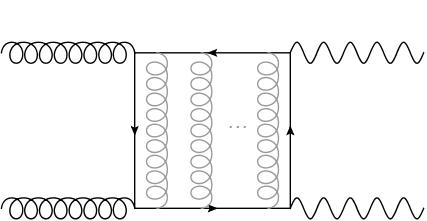}
\caption{}
\end{subfigure}
\caption{Feynman diagrams for $gg \rightarrow \gamma \gamma$ with scalar (a) and fermion (b) loop. Twisted topologies are not shown. Grey gluons indicate ladder diagrams.}
\label{diagram}
\end{figure}

In this Letter we show that any particle charged under $SU(3)_C$ and carrying electric charge  will leave an imprint in the di-photon invariant mass spectrum as it can mediate $gg \to \gamma\gamma$ depicted in Fig.~\ref{diagram}.  
The minimal effect of particle $X$ (we use $X=F$ for a fermion and $X=S$ for a scalar) on di-photon spectrum  depends only on its mass, $M_{X}$, and the combination of its $SU(3)_C$ Dynkin index, $T_{R_X}$, and  electric charge, $Q_X$,  given by
\bea
C_{X}=N_{X} T_{R_X}  Q_X^2,
\label{eq:CX}
\eea
where $N_F$ is the number of Dirac fermions and $N_S$ is the number of complex scalars in case there are more than one particle with the same quantum numbers and similar masses present or the particle is a multiplet under additional symmetry. 

If the particle $X$ is sufficiently long lived to form a narrow bound state then the standard bound state formalism is applicable \cite{Kats:2009bv, Melnikov:1994jb, Fadin:1987wz, Fadin:1988fn, Strassler:1990nw}
and a clear resonance (with ultimate  enhancement factor $\lesssim 100$ for $\Gamma_X \lesssim 10^{-4} M_X$) is expected just below $2M_X$. However this is not the case if the particle $X$ is short-lived which is typical for two-body decays. For example, the lifetime of the top quark ($\Gamma_{\rm top} \simeq 0.8\% m_{\rm top}$) is shorter than toponium formation time. Nevertheless even in this case we show that the effect of particle $X$ can be seen.

For sufficiently large $C_{X}$, the effect appears as a peak-like shape near $2M_X$ with fairly large width as a result of  properties of loop functions, threshold resummation and parton distribution functions (PDFs). 
For smaller $C_X$ the interference with standard model (SM) quarks in the loop of Fig.~\ref{diagram}(b) is important which  results in a dip  around $2M_X$.
Therefore, di-photon searches, especially when the spectrum is measured well with large luminosity, can place strong limits on a variety of new physics, even on scenarios that could otherwise avoid direct detection. Every particle charged under $SU(3)_C$ that have an electric charge leaves its mark in the di-photon spectrum and in principle, with infinite precision, we  could see every one within the energy reach of a given collider.

Current limited data \cite{atlas, cms} only allow us to see new particles with sufficiently large $C_{X}$ and thus, in this Letter, we focus on the region of $C_{X}$ where interference effects are small. We will show the range of  $C_{X}$ as a function of the mass of the particle X already excluded by  measurements of the di-photon spectrum without ever performing direct searches for such particles. We also show future prospects when larger data samples are collected. Finally, we briefly discuss the $C_X$ needed  for a particle with mass around $375$ GeV to explain the excess of events in the di-photon spectrum around 750 GeV.

\noindent
{\bf Glue to light.} 
The individual 1-loop diagrams of a scalar and a fermion  mediating $gg \rightarrow \gamma \gamma$  are shown in black in Fig.~\ref{diagram}.
The gauge invariance  guarantees that the combined amplitude vanishes at $\sqrt{s}=0$, since, as a result of Ward identity, it is proportional to  external momenta. 
Corresponding loop integrals include propagators of the loop particle and they become  enhanced in regions of the phase space that allow the loop particle to be on-shell. When the scattering energy allows two propagators in the loop to be simultaneously on-shell  the amplitude gets significantly enhanced compared with the case below the threshold. This, combined with rapid suppression of gluon PDFs with increasing $\sqrt{s}$,  is the origin of the peak-like shape near $2 M_X$. In addition, resummation of ladder diagrams, shown in grey in Figs.~\ref{diagram}, further enhances the effect near the threshold.

The amplitude for $gg \rightarrow \gamma \gamma$ mediated by particle $X$ can be written as
$A  =  C_X \hat{A}_X$,
where the $\hat{A}_{S(F)}$ is  common for any scalar(fermion); the cross section scales as $C_X^2$ for different $X$.
Our numerical calculation is based on FeynArts, FormCalc and LoopTools \cite{Hahn:2000kx} and we use cteq6l data set for the gluon PDF~\cite{Pumplin:2002vw}. We do not take into account the efficiency in di-photon selection ($\sim$ 50\%) and K-factor from gluon fusion production (expected to be $\sim 1.5$), which do not change the shape of the signal.

\begin{figure}[hbt]
\includegraphics[width=0.46\textwidth]{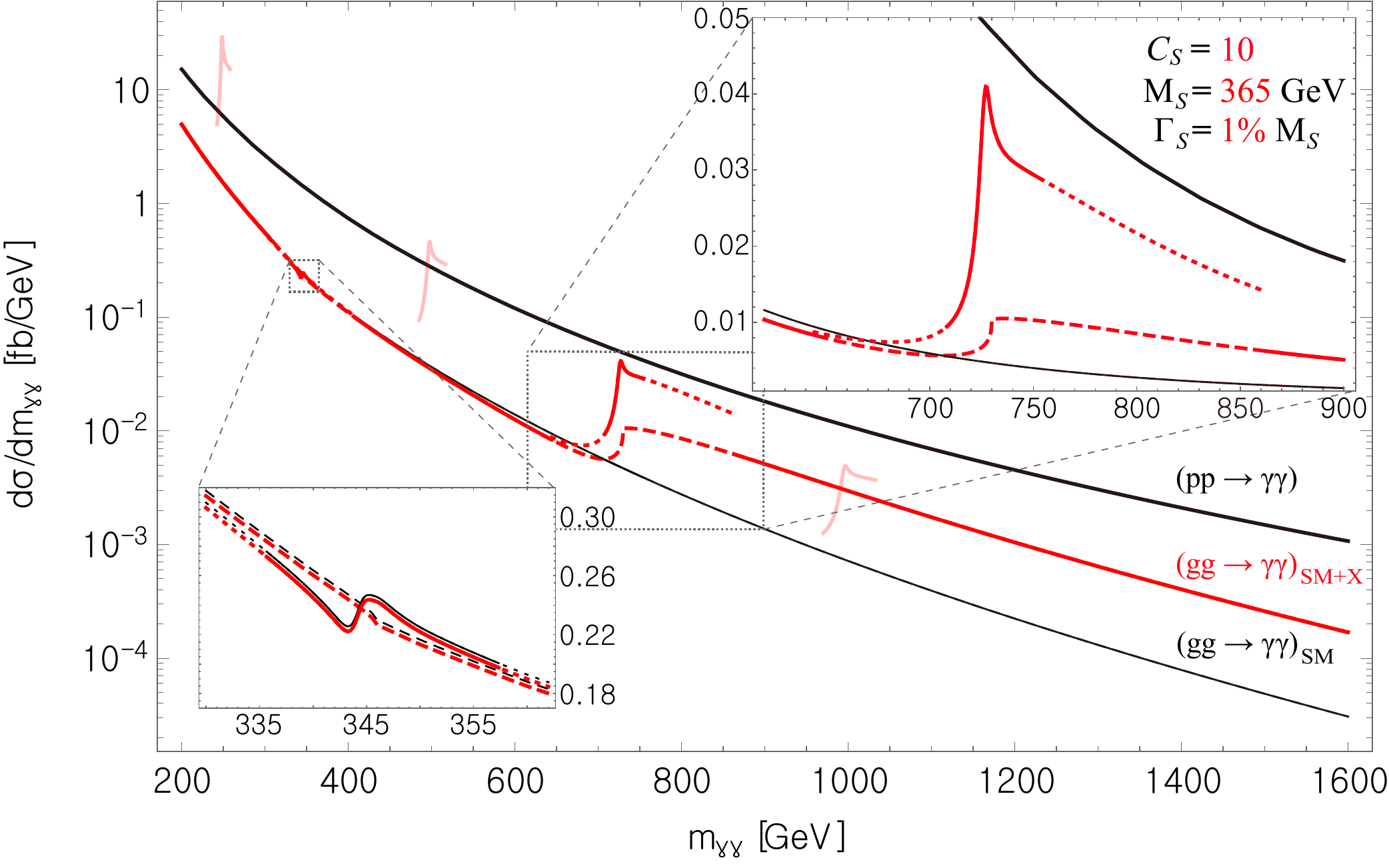}
\includegraphics[width=0.46\textwidth]{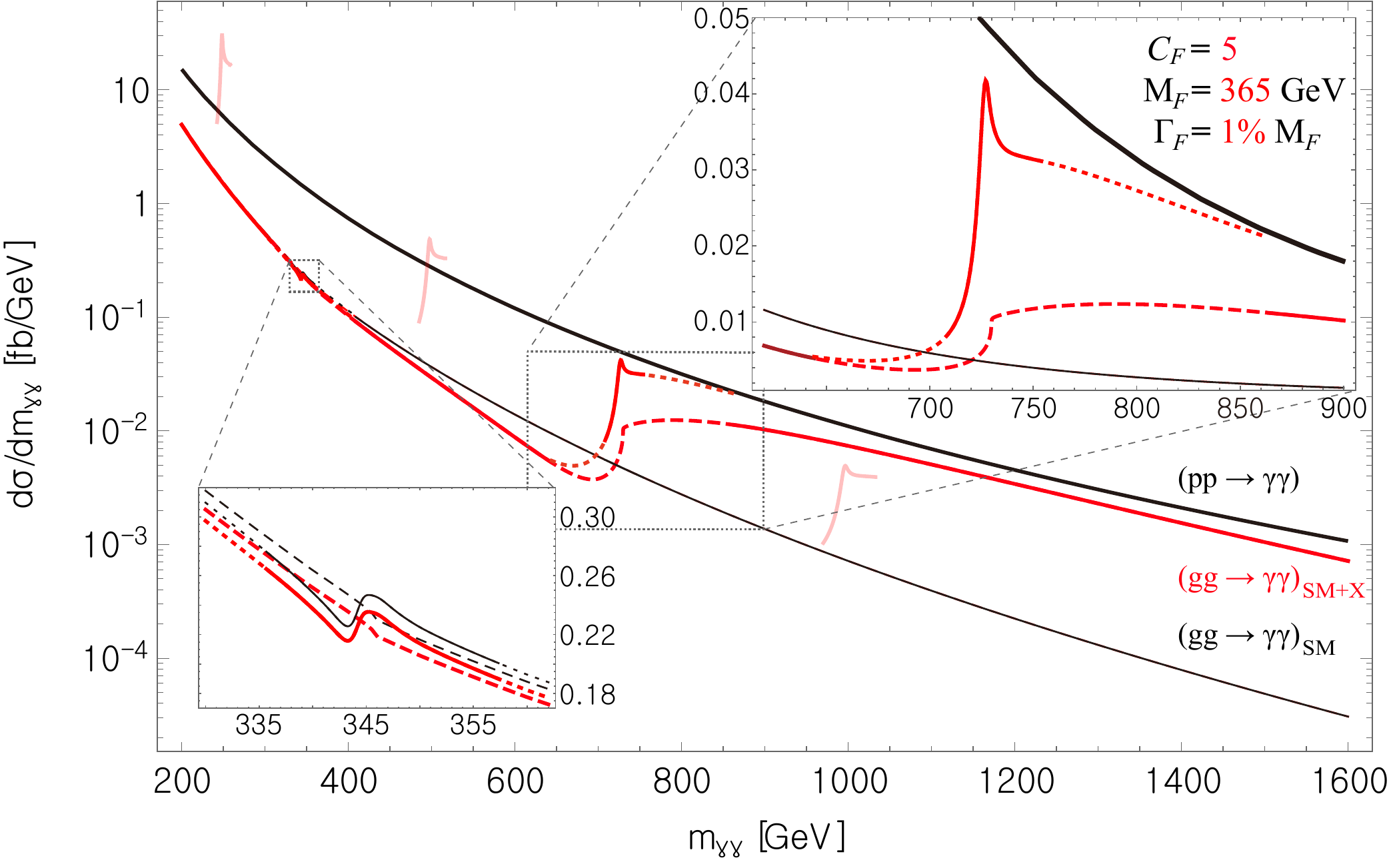}
\caption{The differential cross section for $gg \rightarrow \gamma\gamma$ as a function of di-photon invariant mass $m_{\gamma \gamma}$ for $M_X = 365$ GeV and $\Gamma_X = 1\% M_X$  with  $C_S=10$ for a scalar (top) and  $C_F$=5 for a fermion (bottom) shown in red:  solid lines are 1-loop approximation away from threshold and resummed result near the threshold while dashed and dotted lines are their extensions beyond the valid region. The light red lines are threshold resummed results for  $M_X =$ 125, 250 and 500 GeV. The thin black line is the SM 1-loop result  for $gg \rightarrow \gamma\gamma$. 
The black  line is the fitting function to $pp \rightarrow \gamma\gamma$  from ATLAS~\cite{atlas}.}
\label{xxsection}
\end{figure}

Near the threshold, where $X$ and $ \bar X$  are slowly moving, the ladder diagrams in Fig.~\ref{diagram} are essential since
the $n$ gluon ladder exchanges between $X$ and $ \bar X$ close to on-shell give a factor of $(\bar{\alpha}_s/v)^n$ where $\bar{\alpha}_s$ is the strong coupling at the exchanged gluon momentum scale and $v$ is the velocity of $X$ and $ \bar X$~\cite{Appelquist:1974zd}. The 1-loop amplitude can be  well separated into relativistic and non-relativistic parts near the threshold~\cite{Melnikov:1994jb}. Then the resummed results are obtained by replacing free Green function of $X \bar X$ system by the one which satisfies non-relativistic Schr\"odinger equation of QCD Coulomb potential \cite{Fadin:1987wz, Fadin:1988fn, Strassler:1990nw}. When the electric charge of the particle $X$ is large ($\gtrsim 2$), QED Coulomb potential  should also be taken into account.
 The result of resummation depends on the quadratic Casimir of $X$ and its  decay width $\Gamma_X$. In setting limits, we take a conservative approach and assume $X$ to be the color triplet which gives the smallest quadratic Casimir and we do not include photon ladder exchanges. We also choose $\Gamma_X = 1\% M_X$ as a reference and comment on $\Gamma_X$ dependence of the results.

The differential cross section for $gg \rightarrow \gamma \gamma$ at 13 TeV LHC mediated by SM quarks and the new scalar (fermion) with mass of $365$ GeV and $C_S=10$ ($C_F$=5) is shown with red lines in the top (bottom) plot of Fig.~\ref{xxsection} with the threshold region magnified in the upper right corner. The solid red lines are 1-loop approximation away from threshold and resummed result near the threshold. 
 We also show the  threshold resummed results for different choices of  $M_X$ (light red lines) which indicate that the relative size of the signal of a new particle to the SM background is larger for smaller $M_X$. Finally, the thin black line is the SM 1-loop result  for $gg \rightarrow \gamma\gamma$
and the black  line is the fitting function to $pp \rightarrow \gamma\gamma$  from the recent ATLAS di-photon analysis~\cite{atlas}.

Approaching the threshold,  the 1-loop result cannot be trusted since the next order correction grows as $\bar{\alpha}_s/v$; we nevertheless show it in Fig. \ref{xxsection} in dashed even when larger than 25\% error is expected. On the other hand, the resummed result, which is accurate at the threshold, gets correction proportional to $v$ away from the threshold. When larger than 25 \% error is expected it is only shown as dotted line up to the region where the 1-loop result can be trusted. The two solid red lines are expected to be  smoothly connected once higher order corrections in both calculations are available.

For completeness, the inset in the left bottom corner of Fig.~\ref{xxsection} shows the magnified region near the top threshold, which also illustrates the effect of a particle with small $C_X$ (note, $C_{\rm top }=2/9$). The 1-loop result in the SM, shown in black dashed line, features a kink at the top threshold as a result of the interference with five light quarks~\cite{Dicus:1987fk}. The threshold resummed result (solid lines), which has not been calculated previously,  gives additional  structure.\footnote{ It is noticeable that  presence of a new fermion (red line) causes larger deviation from the SM prediction (black line) compared to a new scalar. This is consistent with the fact that effective 4$\gamma$ operators mediated by a new scalar are one order of magnitude smaller than those mediated by a new fermion~\cite{Fichet:2014uka}.}
 The resulting  dip at $2m_{\rm top }$  is  about 3 \% of the total background. The enhancement from the resummation together with the characteristic shape makes the observation of the top quark in di-photon spectrum more likely in future with high statistics.

\begin{figure}[t]
\includegraphics[width=0.23\textwidth]{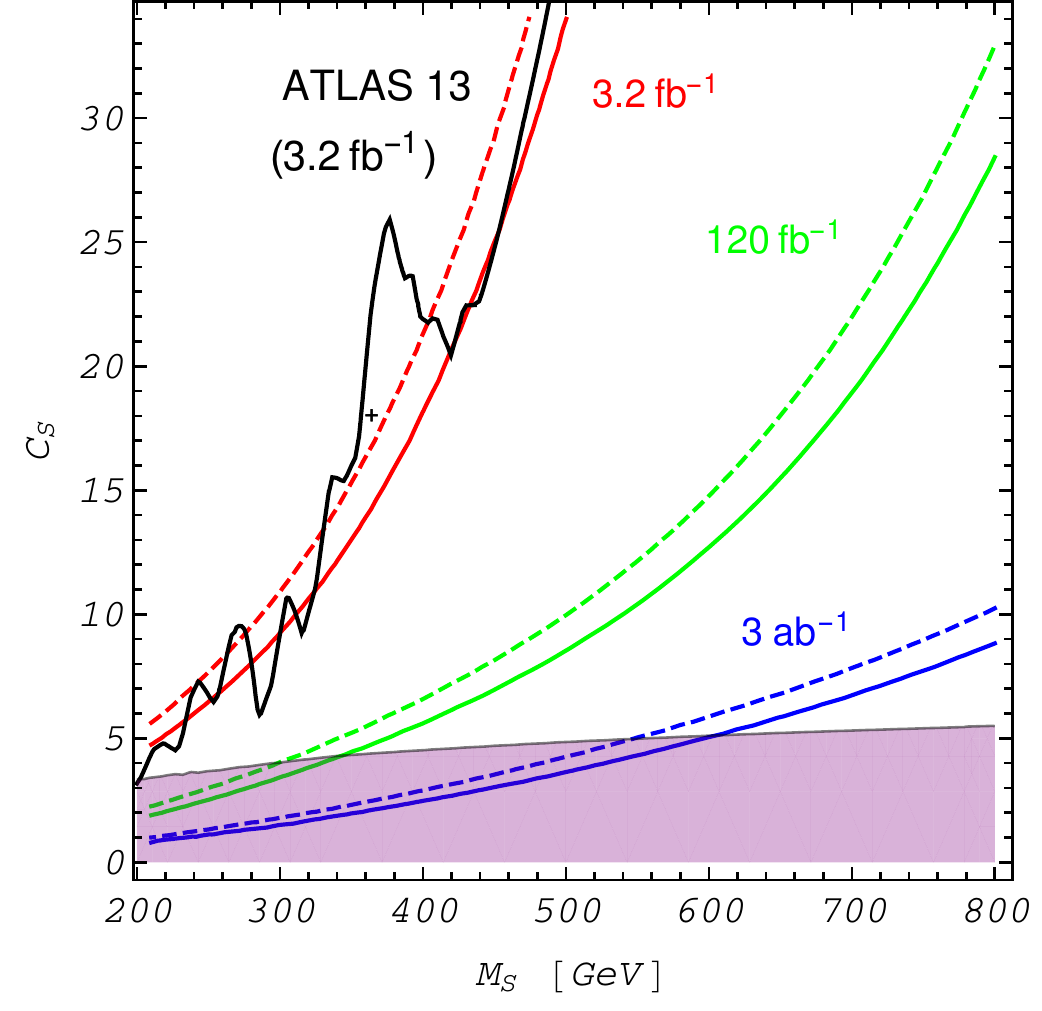}
\includegraphics[width=0.23\textwidth]{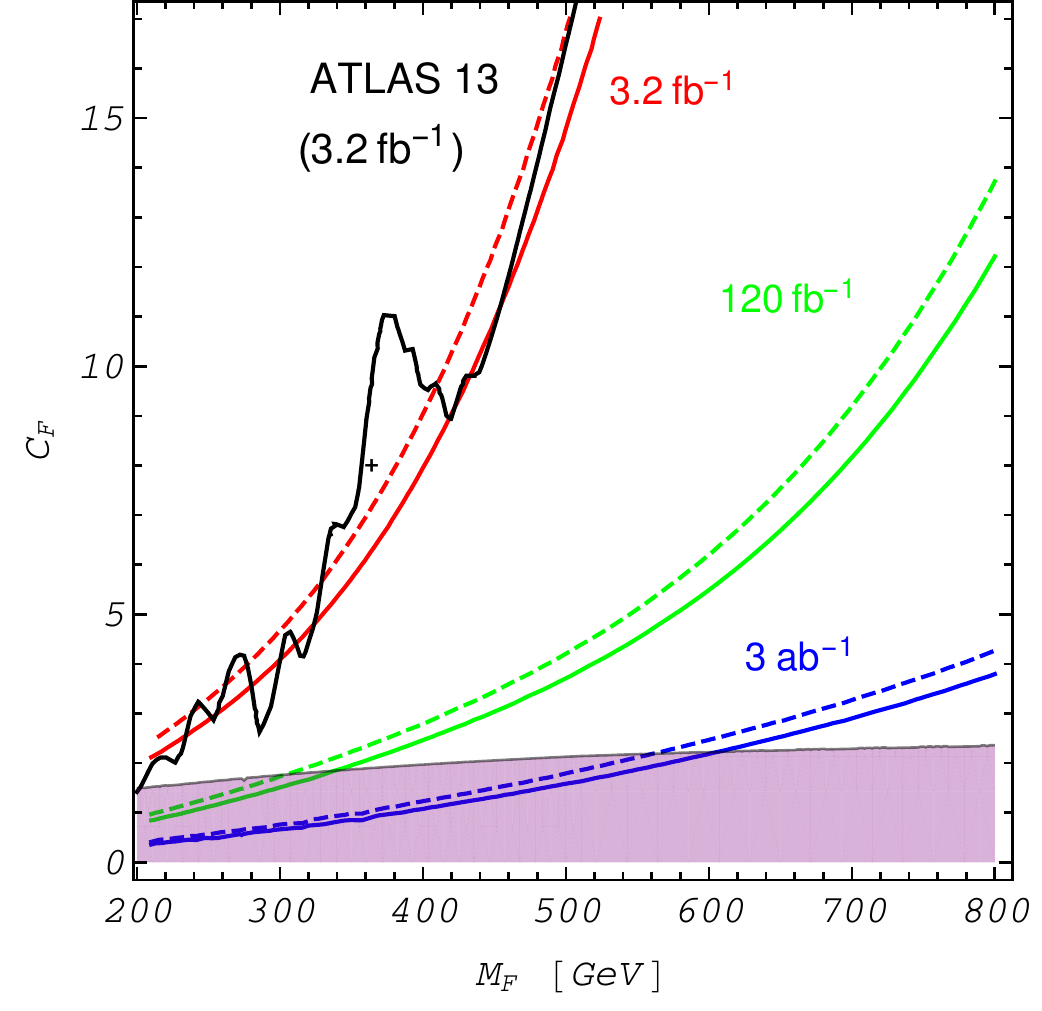}
\caption{The 95\% C.L. upper limits (black) and expected upper limits (solid red) on $C_S$ (left) and $C_F$ (right) assuming $\Gamma_X=1\% M_X$ based on ATLAS analysis with current integrated luminosity of 3.2 ${\rm fb}^{-1}$ at  13 TeV LHC. 
The green and blue solid lines correspond to projections for integrated luminosity of 12 ${\rm fb}^{-1}$ and 3 ${\rm ab}^{-1}$ assuming $\Gamma_X=1\% M_X$.
Corresponding dashed lines are expected upper limits for $\Gamma_X=3\% M_X$.
}
\label{contour}
\end{figure}

\noindent
{\bf Results.} From current data, we can set the upper limit on $C_X$  by comparing the expected signal for given $M_X$ with the background.
In Fig.~\ref{contour}, the  black line (solid red line) indicates the $95\%$ C.L. upper limit (expected upper limits) on $C_X$  as  a function of $M_X$ assuming $\Gamma_X = 1\% M_X$ 
from current ATLAS data corresponding  to the integrated luminosity of 3.2 ${\rm fb}^{-1}$ at 13 TeV LHC. For the expected upper limits, 
the statistical significance, $S/\sqrt{B}$, is calculated with background given by the data-driven fitted function of the ATLAS analysis~\cite{atlas}.
The signal is selected only from the  region in which the threshold resummation is trustable, as is shown with solid lines near the threshold in Fig.~\ref{xxsection}, which roughly correspond to $\pm 3\%$ around $2M_X$.
 We also take a simple approach based on the number of expected events and the information about the shape of the signal is not used. Thus, the resulting exclusion limits are conservative.  
 
 The obtained limits are valid for any $\Gamma_X < 1\% M_X$. Actually, as the width decreases, 
  the exclusion limits are stronger due to the enhancement factor corresponding to the formation of the bound state mentioned in the introduction.
 For larger width, $\Gamma_X = 3\% M_X$,  the expected  upper limits (dashed red line) are slightly weaker. Note, most of models with perturbative couplings have particles with widths smaller than 3\% of their masses. 

The green and blue solid (dashed) lines in Fig.~\ref{contour} correspond to projections for integrated luminosity of  12 ${\rm fb}^{-1}$ (expected for Run 2) and 3 ${\rm ab}^{-1}$ (for HL-LHC) with $\Gamma_X=1\%M_X$  (3\%$M_X$). 
The shaded ranges in Fig.~\ref{contour} indicate values of $C_X$ for which the SM amplitude is of the same size or larger than the amplitude with $X$ in the loop. For smaller $C_X$ the peak-like shape  is gradually replaced by a dip around $2M_X$, similar to what we saw for the top quark in Fig.~\ref{xxsection}.  The overall signal strength remains significant even for smaller $C_X$. Compared to the top threshold which interferes with 5 light quarks, $X$ interferes with 6 quarks, further amplifying the effect. However, in this case a more sophisticated analysis based on the shape of the signal is required which is beyond the scope of this Letter. The dependance of the shape of the signal on $C_X$ and $\Gamma_X$ will be presented elsewhere.

 As we can see from Fig. \ref{contour}, di-photon searches already place limits on possible  new particles irrespectively of their decay modes. For example, a 200 GeV scalar particle with $C_S \gtrsim  3$ or a fermion with $C_F \gtrsim 1.5$ are already excluded for $\Gamma_X \le 1\% M_X$. The latter corresponds to, for example, a triplet under $SU(3)_C$ with $NQ^2 \gtrsim 3$ (for example, one vector-like quark with electric charge $5/3$ gives $NQ^2=2.8$)
or an octet with $NQ^2 \gtrsim 0.5$.

Let us discuss the possibility that the excess of events in the di-photon invariant mass spectrum around 750 GeV originates from $gg\to \gamma\gamma$ mediated by loops of a particle with mass around 375 GeV.

\begin{figure}[t]
\includegraphics[width=0.46\textwidth]{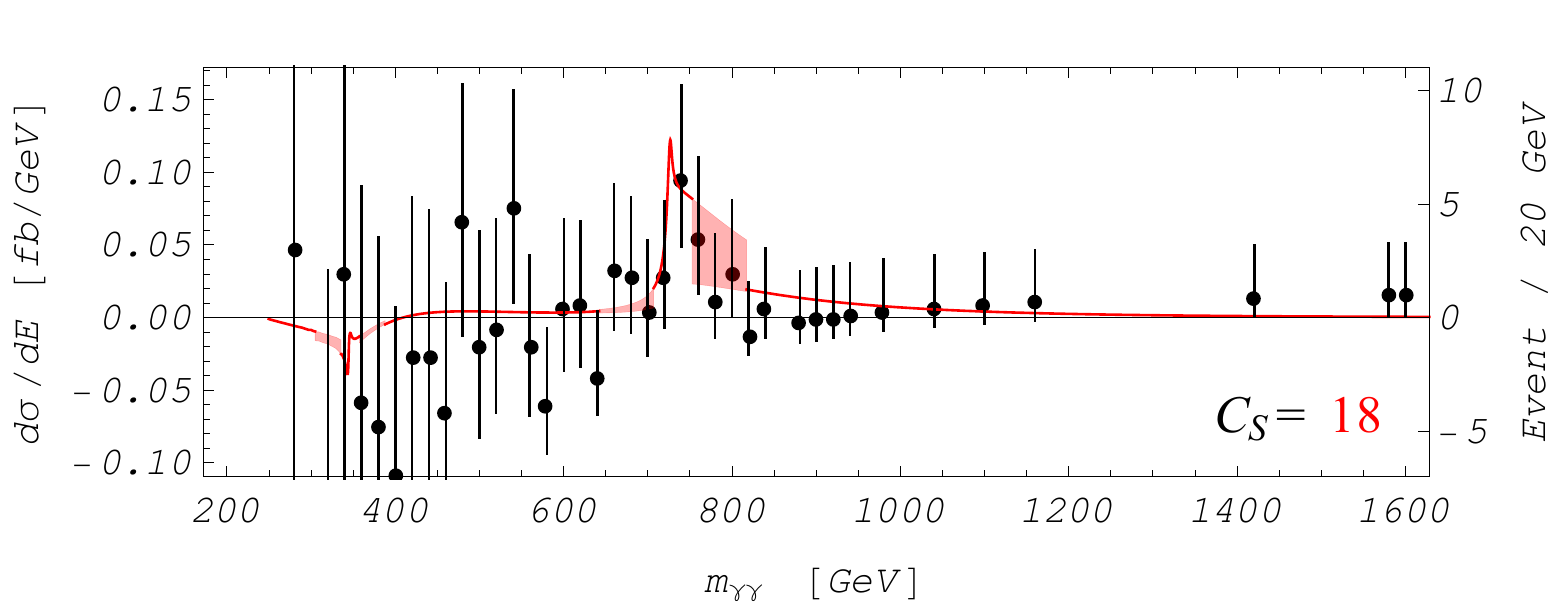}
\includegraphics[width=0.46\textwidth]{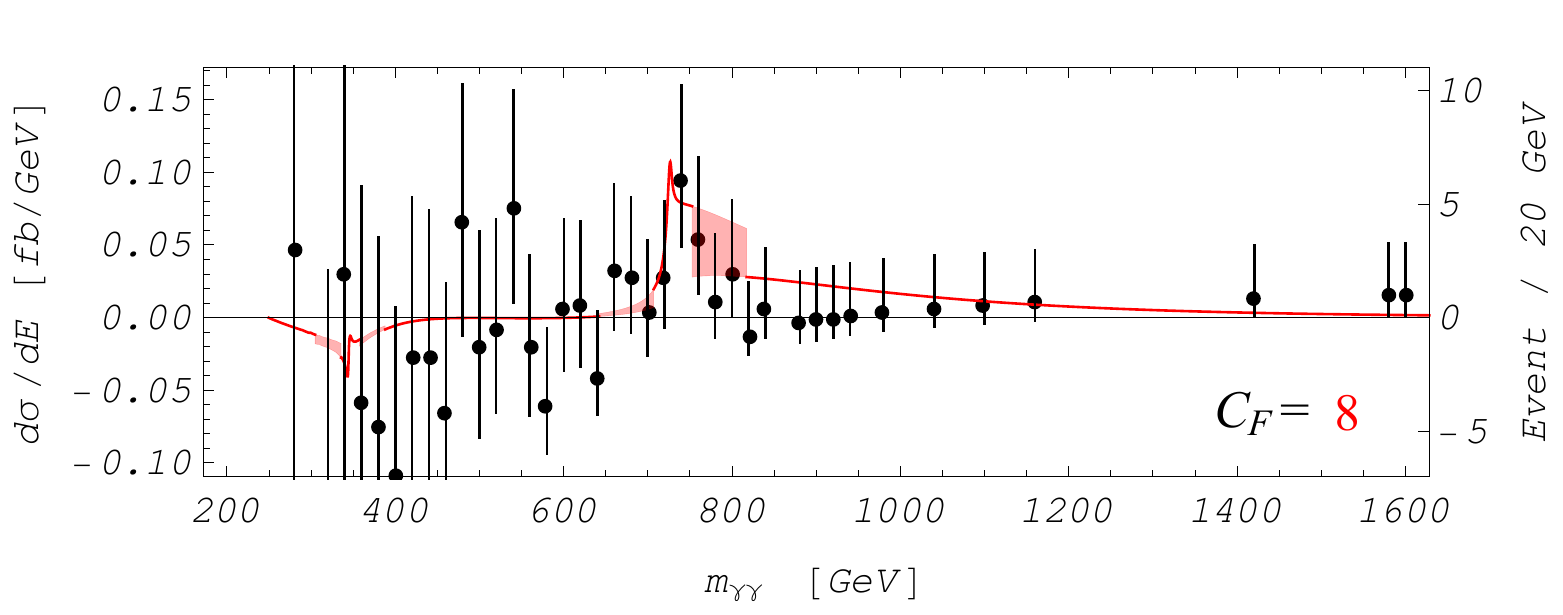}
\caption{
Contribution of a  scalar (top) and a fermion (bottom) with $M_X = 365$ GeV and $\Gamma_X=1\%M_X$ to the di-photon spectrum for $C_S=18$ ($C_F=8$) in the data minus the fitted function plot from ATLAS~\cite{atlas}. The meaning of solid lines is the same as in  Fig.~\ref{xxsection} and the shaded area extends from the 1-loop result to the resummed result in the region where higher order corrections are needed to connect solid lines.
The dip from the top quark is also visible at $2m_{\rm top }$.}
\label{750excess}
\end{figure}

In Fig. \ref{750excess}, we show the contribution of a 365 GeV scalar particle (top) and a fermion (bottom) to the di-photon spectrum for $C_S=18$ ($C_F=8$) in the data minus the fitted function plot from the ATLAS analysis~\cite{atlas}. The meaning of solid lines is the same as in  Fig.~\ref{xxsection} and the shaded area extends from the 1-loop result to the resummed result in the region where higher order corrections are needed to connect solid lines (indicated by dashed and dotted lines in Fig.~\ref{xxsection}). In order to illuminate our previous discussion we also show the effect of the top quark - the dip at $2m_{\rm top }$. The parameter choices of Fig. \ref{750excess} are also marked in  Fig.~\ref{contour}. The scalar case is more consistent with data points and if the preferred large width (45 GeV) of the excess persists with more data, this explanation would be especially well motivated.\footnote{For another possible explanation of  the excess  mimicking a particle with a large width see Ref.~\cite{Cho:2015nxy}.}

A particle with such a large $C_X$ may seem highly unusual but its existence is not necessarily ruled out by direct searches. Consider for example a vector-like quark that couples to a light SM quark and a SM singlet scalar $S$ (large $C_X$ can be achieved by large $N$ in Eq.~(\ref{eq:CX})).  Although $X$ can be pair-produced at high rates, it decays into $S$+jet. The $S$ can further cascade decay through other SM singlets and finally the lightest one can decay back to SM light quarks through a coupling originating from small mixing with the SM Higgs boson, as in models with singlet extensions of the Higgs sector~\cite{Dermisek:2009si}. The signature of the pair produced $X\bar X$ is thus a (possibly large) number of jets.\footnote{ 
Alternatively, instead of large $N$ one might consider an $X$ with large hypercharge. In this case $S$ has to carry hypercharge as well, but the rest of the discussion above holds with small modifications and the same conclusions.} Currently any scenario with pair-produced particles each decaying to more than 4 jets is essentially unconstrained. In addition, masses of SM singlets could be such that some of the final state jets are typically significantly softer than others, making the mass reconstruction of $X$ challenging or impossible even  with large data sets in future.\footnote{For a scalar $X$, the $S$ has to be a SM singlet fermion, it can decay into other SM singlet scalars and fermions, eventually decaying into jets and one stable SM singlet fermion leading to, depending on the masses,  small amount of missing energy in  final states, again resulting in the same conclusions.}

There are endless variations of these scenarios. Designing direct searches for all these and other, perhaps better motivated, hard-to-see scenarios might be difficult or not possible. 
However, precise di-photon spectrum measurements can lead to discoveries or strong constraints irrespectively of prospects for direct detection of new particles.

\noindent
{\bf Acknowledgements}
We thank Sunghoon Jung, Yeo Woong Yoon and especially Yevgeny Kats for discussion.
This work was supported by the National Research Foundation of Korea (NRF), No. 0426-20140009 and No. 0409-20150110. The work of RD was supported in part by the U.S. Department of Energy under grant number {DE}-SC0010120
and by  the Ministry of Science, ICT and Planning (MSIP), South Korea, through the Brain Pool Program.


\begin{thebibliography}{00}


  
\bibitem{Fadin:1987wz} 
  V.~S.~Fadin and V.~A.~Khoze,
  JETP Lett.\  {\bf 46}, 525 (1987)
  [Pisma Zh.\ Eksp.\ Teor.\ Fiz.\  {\bf 46}, 417 (1987)].
  
\bibitem{Fadin:1988fn} 
  V.~S.~Fadin and V.~A.~Khoze,
  Sov.\ J.\ Nucl.\ Phys.\  {\bf 48}, 309 (1988)
  [Yad.\ Fiz.\  {\bf 48}, 487 (1988)].
    
\bibitem{Strassler:1990nw} 
  M.~J.~Strassler and M.~E.~Peskin,
  Phys.\ Rev.\ D {\bf 43}, 1500 (1991).
  doi:10.1103/PhysRevD.43.1500


\bibitem{Melnikov:1994jb} 
  K.~Melnikov, M.~Spira and O.~I.~Yakovlev,
  Z.\ Phys.\ C {\bf 64}, 401 (1994)
  doi:10.1007/BF01560100
  [hep-ph/9405301].
  

\bibitem{Kats:2009bv} 
  Y.~Kats and M.~D.~Schwartz,
  JHEP {\bf 1004}, 016 (2010)
  doi:10.1007/JHEP04(2010)016
  [arXiv:0912.0526 [hep-ph]].





\bibitem{atlas} 
  The ATLAS collaboration,
  ATLAS-CONF-2016-018.


\bibitem{cms}
The CMS Collaboration,  CMS-PASEXO-15-004.


\bibitem{Hahn:2000kx} 
  T.~Hahn,
  Comput.\ Phys.\ Commun.\  {\bf 140}, 418 (2001)
  [hep-ph/0012260].
  T.~Hahn and C.~Schappacher,
  Comput.\ Phys.\ Commun.\  {\bf 143}, 54 (2002)
  [hep-ph/0105349].
  T.~Hahn and M.~Perez-Victoria,
  Comput.\ Phys.\ Commun.\  {\bf 118}, 153 (1999)
  [hep-ph/9807565].
  T.~Hahn and C.~Schappacher,
  Comput.\ Phys.\ Commun.\  {\bf 143}, 54 (2002)
  [hep-ph/0105349].



\bibitem{Pumplin:2002vw} 
  J.~Pumplin, D.~R.~Stump, J.~Huston, H.~L.~Lai, P.~M.~Nadolsky and W.~K.~Tung,
  JHEP {\bf 0207}, 012 (2002)
  [hep-ph/0201195].

\bibitem{Appelquist:1974zd} 
  T.~Appelquist and H.~D.~Politzer,
  Phys.\ Rev.\ Lett.\  {\bf 34}, 43 (1975).
  doi:10.1103/PhysRevLett.34.43
  
\bibitem{Dicus:1987fk} 
  D.~A.~Dicus and S.~S.~D.~Willenbrock,
  Phys.\ Rev.\ D {\bf 37}, 1801 (1988).
  doi:10.1103/PhysRevD.37.1801


  

\bibitem{Fichet:2014uka} 
  S.~Fichet, G.~von Gersdorff, B.~Lenzi, C.~Royon and M.~Saimpert,
  JHEP {\bf 1502}, 165 (2015)
  doi:10.1007/JHEP02(2015)165
  [arXiv:1411.6629 [hep-ph]].
  
\bibitem{Cho:2015nxy} 
  W.~S.~Cho, D.~Kim, K.~Kong, S.~H.~Lim, K.~T.~Matchev, J.~C.~Park and M.~Park,
  Phys.\ Rev.\ Lett.\  {\bf 116}, no. 15, 151805 (2016)
  doi:10.1103/PhysRevLett.116.151805
  [arXiv:1512.06824 [hep-ph]].

\bibitem{Dermisek:2009si}
Similar scenarios were considered in  R.~Dermisek,
  Mod.\ Phys.\ Lett.\  A {\bf 24}, 1631 (2009)
  [arXiv:0907.0297 [hep-ph]].







  
\end{thebibliography}
\end{document}